\documentclass[conference]{IEEEtran}
\IEEEoverridecommandlockouts
\usepackage{cite}
\usepackage{amsmath,amssymb,amsfonts}
\usepackage{algorithmic}
\usepackage{graphicx}
\usepackage{textcomp}
\usepackage{xcolor}
\usepackage{url}
\def\BibTeX{{\rm B\kern-.05em{\sc i\kern-.025em b}\kern-.08em
    T\kern-.1667em\lower.7ex\hbox{E}\kern-.125emX}}
\begin{document}

\title{An Investigation of the AUTOSAR Adaptive Platform from an Industry Perspective\\

\thanks{The authors would like to thank TRATON R\&D AB, Software Center, and VINNOVA. This study was partially funded by Software Center, an initiative of Chalmers, University of Gothenburg, 17 companies, and two universities.}
}
\newif\ifblind
\blindfalse        

\ifblind
  \author{%
    \IEEEauthorblockN{Anonymous Authors}
    \IEEEauthorblockA{Paper under double-blind review}
  }
\else
\author{
\IEEEauthorblockN{1\textsuperscript{st} Bengt Haraldsson}
\IEEEauthorblockA{
\textit{TRATON R\&D AB, Södertälje}\\
\textit{Chalmers University of Technology}\\
\textit{University of Gothenburg}\\
Gothenburg, Sweden\\
bengthar@chalmers.se}
\and
\IEEEauthorblockN{2\textsuperscript{nd} Srijita Basu \\ 3\textsuperscript{rd} Miroslaw Staron}
\IEEEauthorblockA{
\textit{Chalmers University of Technology}\\
\textit{University of Gothenburg}\\
Gothenburg, Sweden\\
srijita.basu@gu.se / miroslaw.staron@gu.se}
\and
\IEEEauthorblockN{4\textsuperscript{th} Erika Mayer}
\IEEEauthorblockA{
\textit{TRATON R\&D AB}\\
\textit{Extra Consulting}\\
Södertälje, Sweden\\
erika.mayer@scania.com}
}
\fi
\maketitle

\begin{abstract}
\textbf{Context}: The reliance on software as a distinguishing factor in the automotive industry is increasing. With a combined reliance on vendor-supplied software and cost-effective implementation, the AUTOSAR consortium was initialized to provide standardized platform specifications that enable re-use. Specifically, the AUTOSAR Adaptive Platform (AP) specification aims to provide a high-performance service-oriented architecture. \textbf{Objective}: The goal of this study is to investigate what pain-points emerge when developing AUTOSAR Adaptive applications and whether they originate from the platform specification, its vendor-implementation, or its local usage. \textbf{Methods}: We conduct a Design Science Research study, developing a minimal AP that serves as an experimental prototype for our investigation. \textbf{Results}: We find that a combination of specification-inherent, implementation-based, and local practices contributes to the emergence of pain-points. \textbf{Conclusions}: We conclude that there are AUTOSAR specification-inherent reasons for pain-points, resulting from architectural choices and re-use goals. The implication for development organizations is the need to mitigate these effects through tooling that better supports configuration file management and reduces developer training time to properly understand the adaptive application runtime life-cycle.
\end{abstract}


\section{Introduction}
\label{sec:intro}
In the current landscape of automotive products, software has become an increasingly important distinguishing factor driving business success or failure \cite{staron2021automotive}. With increased competition from new entrants into markets offering innovative software-based solutions, traditional Original Equipment Manufacturers (OEM) and their suppliers need to find ways to compete. 

In order to promote effective, efficient, safe, and secure software development the AUTOSAR Classical and AUTOSAR Adaptive Platform (AP) aim to create an open industry standard for automotive industry E/E architecture \cite{AUTO}. The first platform version for traditional real-time embedded systems development was released in 2005 (Classical) and the first version for service oriented architectures in 2017 (Adaptive). We have good understanding of the technical implications of the Adaptive framework from the descriptions in the specifications, but the organizational implications in development organizations is less understood.

This study aims to investigate the AP from an industry perspective. We want to know what happens in an organization when a standardized architecture framework that is specified to meet a general set of requirements is introduced in a local development context. 

Our investigation is based around the notion of \textit{pain-points}. Inspired by Chattopadhyay et al. \cite{Chattopadhyay}, we define the pain-points associated with AP development as; \emph{Developer-experienced difficulties (cognitive, technical, or organizational) that emerge when applying the AUTOSAR Adaptive architectural framework within safety-critical, large-scale automotive product development.} In this study, pain-points refer to recurring and structurally rooted difficulties rather than incidental or individual frustrations.

We associate pain-point emergence with increase of two different classes of complexity. The two classes are; 1) system internal code complexity (e.g., due to code complexity increase from local adaption of the general framework), or 2) system external complexity (e.g., due to workarounds needed in vendor provided tooling to accommodate local process requirements) \cite{Antinyan_revealing}. However, due to the tight coupling of local processes, ways of working, vendor supplied software, and the AP specification itself, root causes of pain-points are obfuscated and not easily pinpointed in the local context. 

Thus, we wanted to investigate locally emergent phenomenon, but in the local context it's not clearly observable how the phenomenon occurs. To alleviate this, we conducted a Design Science Research project together with an industrial partner. Industry data was collected from TRATON R\&D AB, a European development organization that supplies original equipment manufacturers (OEMs) with hardware and software solutions. There, we developed a minimal AP-stack as a driver for experimental evaluation of the AP design specifications. We first collected qualitative data from a series of workshops to find a classification taxonomy of pain-points. We then build a minimal (prototype) AP based on the AUTOSAR specifications \cite{AUTO}, evaluated our prototype, and had two developers perform experimental application development on the platform. After the experimental development we collected further developer experience qualitative data from workshops to find root causes and possible technical debt.

Pain-points negatively affect Developer Experience (DevEx), which in turn influences productivity and software quality~\cite{Palomino,Fagerholm,Graziotin}. Understanding their emergence is therefore critical for organizations developing software-intensive products. For this reason we conduct our study guided by the following research questions:

\begin{itemize}
    \item \textbf{RQ1:} What are the possible pain-points related to system development on an AUTOSAR Adaptive platform?
    \item \textbf{RQ2:} Are the reasons for the presence of these pain-points inherent in the complexities introduced by the architecture proposed by AUTOSAR, and if so, what are the root causes?
    \item \textbf{RQ3:} What possible negative impacts are likely to arise from these pain-points and complexities?
\end{itemize}

This paper contributes 1) a taxonomy of developer pain-points in AUTOSAR Adaptive, 2) a diagnostic minimal stack to isolate specification- from implementation-induced complexity, and 3) an analysis of specification-inherent versus contextual pain-points and their organizational implications.

The rest of this paper is structured as follows. Section \ref{sec:background} introduces the main concepts and structure of the AUTOSAR Adaptive platform and related work in the area. In Section \ref{sec:method}, we present the methodology of the study, and Section \ref{sec:results} summarizes our results. We discuss the results and validity of our study in Section \ref{sec:discussion}. Finally, Section \ref{sec:conclusion} concludes the paper.

\section{Background and Related Work}
\label{sec:background}
The AUTOSAR Adaptive Platform \cite{AUTO} is an industry standard developed within the AUTOSAR consortium to address the increasing complexity of software in modern vehicles. While the AUTOSAR Classic Platform has long been used in embedded, real-time control applications (e.g., engine control units) \cite{staron2021automotive}, the Adaptive Platform was introduced to support new classes of applications such as advanced driver assistance systems (ADAS), autonomous driving, and connected services \cite{AUTO}. These applications demand significantly higher computational power, dynamic software updates, and communication with off-board cloud services.

The main driver for the creation of the AUTOSAR platforms (Classical and Adaptive) has been the distributed development of software systems where OEMs integrate software and hardware from TIER1, TIER2, and TIER3 level suppliers \cite{staron2021automotive}. To reduce cost of parts it was important to establish a standard where suppliers could re-use software developed for one OEM when starting new development projects targeting another OEM. However, the complexity introduced by changes to architecture in this context is hard to measure due to this "black-box" situation in the integrated solution (i.e., on complete vehicle level) \cite{Durisic}.

The Adaptive Platform differs fundamentally from Classic in its target environment and programming model \cite{AUTO}. It is designed for high-performance computing units (HPCs) running general-purpose operating systems such as Linux, QNX, or POSIX-compliant RTOSes \cite{staron2021automotive}. Applications are developed in C++ and deployed as processes rather than statically configured tasks \cite{AUTO}. This enables dynamic deployment, service discovery, and update-over-the-air scenarios, which are central to current and future vehicle software architectures.

\begin{figure}
    \centering
    \includegraphics[width=0.7\linewidth]{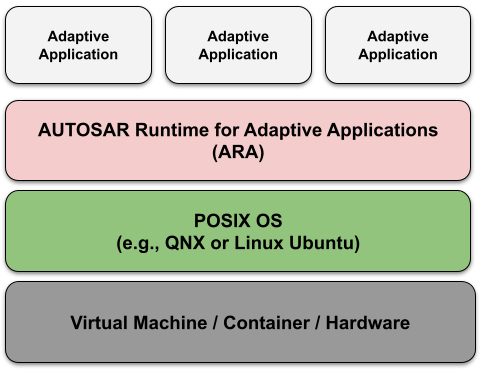}
    \caption{Simplified visualization of adaptive applications running on ARA and the layers below.}
    \label{fig:ARA_structure}
\end{figure}

Applications running on the AP are developed as one or several Software Clusters (SWCs) that run as separate processes on the Runtime for Adaptive Applications (ARA) (see Figure \ref{fig:ARA_structure}) \cite{PlatformDesign}. For example, an application may compare \texttt{set\_speed} and \texttt{current\_speed} and emit a warning \texttt{Event}. Such applications depend on multiple services and subscriptions, introducing dependency and communication complexity.

\begin{figure}
    \centering
    \includegraphics[width=0.9\linewidth]{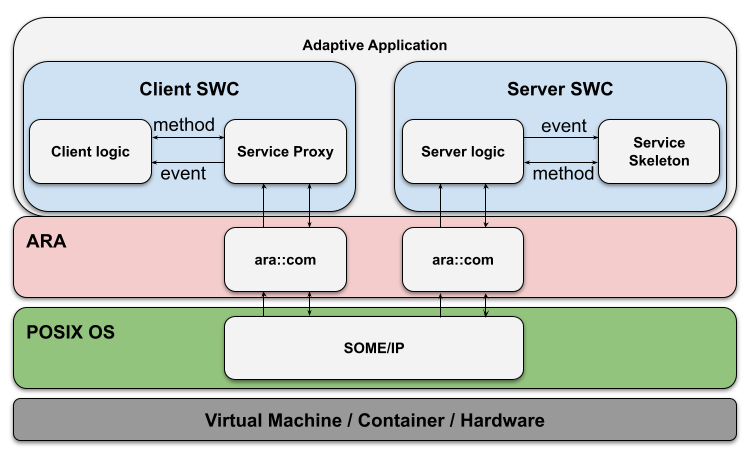}
    \caption{Client--server communication. Illustrated with SOME/IP since this is recommended by Autosar. Note that it is not necessarily the case that the Client SWC and Server SWC runs on the same Hardware. Here depicted so to simplify the illustration.}
    \label{fig:client_server}
\end{figure}

Key architectural elements of ARA as described in \cite{PlatformDesign} include:

\textit{Service-Oriented Communication (SoC) Management:} Applications interact using service discovery and communication mechanisms, typically over SOME/IP (recommended by AUTOSAR \cite{PlatformDesign}) or DDS bindings (see Figure \ref{fig:client_server}). This design allows applications to be distributed across ECUs or integrated with external cloud services, but also require specification in AUTOSAR XML (ARXML) files that can be long and intricate to traverse. The ARXML files are normally generated from tooling.

\textit{Execution Management} handles application lifecycle, while \textit{Platform Services} provide APIs for logging, diagnostics, and persistent storage. \textit{Safety and security} considerations introduce additional constraints and trade-offs in the Adaptive architecture \cite{PlatformDesign}. 

By providing these standardized interfaces and runtime services, the Adaptive Platform aims to decouple application development from underlying hardware and supplier implementations \cite{AUTO}. 

\subsection{Related Work}
There are specific investigations relating to the AUTOSAR Adaptive specifications. For example, Menard et al. investigated determinism in the Adaptive Platform \cite{Menard}. They identified three possible sources of non-determinism in the platform specification: 1) The suggested SWC programming model is to base them on threads; 2) the undefined order of processing for incoming messages to a SWC; and 3) the lack of a formal requirement to achieve point-to-point in-order messaging, likely due to the fact that the duration of message transfer cannot be guaranteed. The first source of non-determinism could be considered solved by the \textit{deterministic client} specification \cite{Menard}. However, these requirements have been removed in later releases of the Adaptive Platform specifications (see e.g., \cite{ExecMgmt}). It's clear that special considerations will need to be taken when determinism is crucial. Another example is Bhat et al. who designed a framework for the fault-tolerant execution of applications on the AP \cite{Bhat} and found some gaps in the AP 18.10 standard.

Some authors have investigated co-operability of AUTOSAR Adaptive and other platforms or standards. For example, Hong \& Moon propose an architecture for interoperability between the Robotic Operating System 2 (ROS2) and the Adaptive Platform to bridge the gap between autonomous drive experimentation---often carried out on ROS2---and industrial practice in the automotive industry \cite{Hong}. 

Helmy et al. implemented an integration of the Service Oriented Vehicle Diagnostics (SOVD) with AP \cite{Helmy}. They found that this simplifies the implementation of HPC diagnostics.

Related to the complexity from communication dependencies between services and applications over SOME/IP, Durisic et al. introduce a measure for coupling and cohesion complexity in automotive systems that aligns with a hierarchical and sub-contractor-based development methodology. They found that change management processes based on dependency analysis are crucial in the automotive systems domain \cite{Durisic}. This relates to our context because of the way AP SWCs and Applications communicate and the high degree of dependencies that need to be specified in the ARXML-files. 

Although related, there is no comprehensive work that answers our questions regarding the impact of introducing AUTOSAR Adaptive in a local context, leaving us confident of our study's novelty.

\section{Methodology}
\label{sec:method}
\begin{figure}
    \centering
    \includegraphics[width=0.99\linewidth]{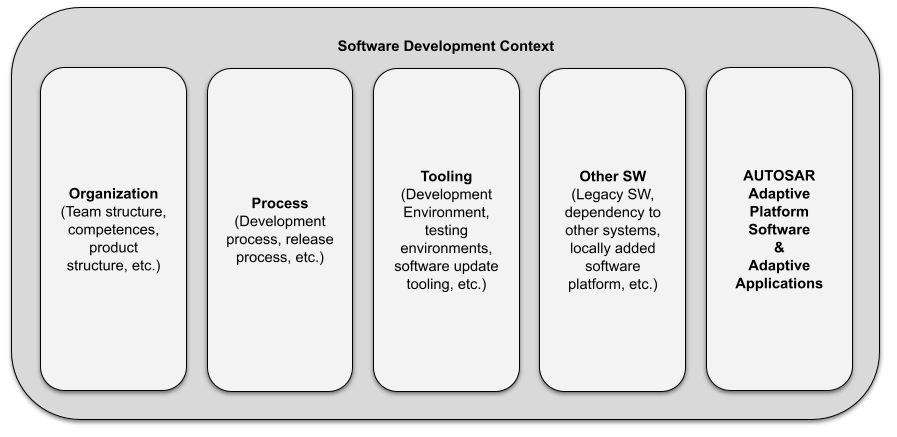}
    \caption{Showing influences on emergent pain-points in a software development context. For illustration simplicity these have been separated. However, in the local context the distinction became blurred and where one influencing factor ended and another began became hard to pinpoint.}
    \label{fig:AA_influence}
\end{figure}

\begin{figure}
    \centering
    \includegraphics[width=0.99\linewidth]{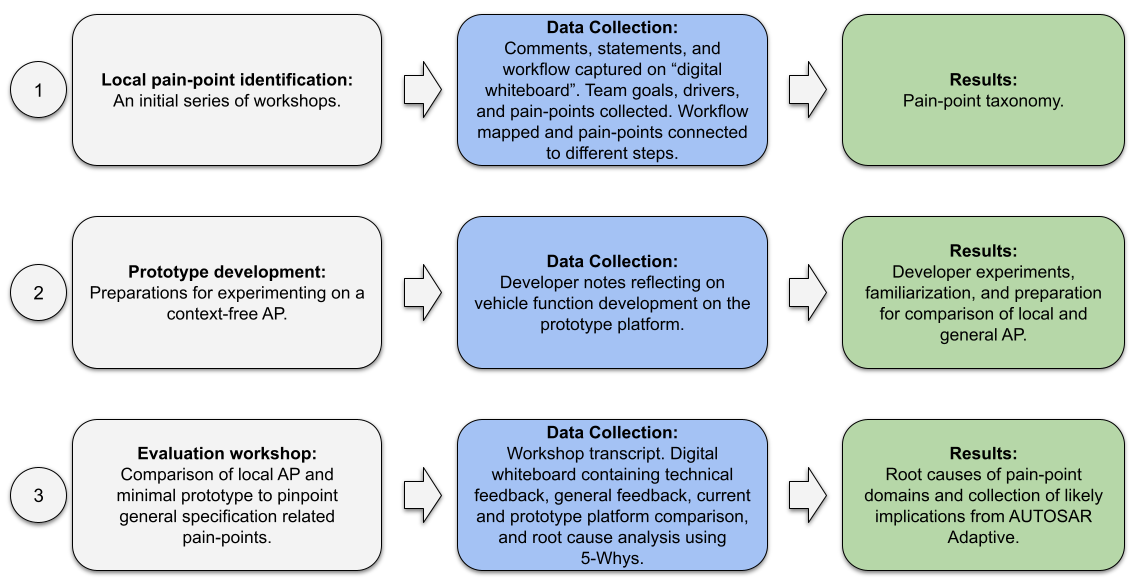}
    \caption{Showing the steps employed in our study process to extract a taxonomy from local pain-points, assessing the root causes of pain-point domains, and likely negative impacts from using AUTOSAR Adaptive in a development organization.}
    \label{fig:process_steps}
\end{figure}

In order to investigate the origins and consequences of pain-points (see Section \ref{sec:intro} for a definition) experienced by developers using a software platform based on the AP, we conduct a Design Science Research (DSR) study embedded in an industrial context \cite{Hevner}. We largely follow the steps described in \cite{Peffers}, building a prototype (minimal for our experiments) of the AP in order to isolate the influence of other factors' on identified pain-points. Figure \ref{fig:AA_influence} shows a depiction of factors that all affect the emergence of pain-points and who's influence are hard to distinguish in the local context via observation alone. DSR is appropriate for this work, as it supports the creation and evaluation of artifacts (our minimal AP) that contribute both to practical problem-solving and theoretical understanding. Figure \ref{fig:process_steps} shows the three-step process applied in this study.

\subsection{Industrial Context Description}
TRATON R\&D AB supplies hardware and software systems to OEMs within the heavy vehicle segment owned by the TRATON Group. Traditionally, the company has historically developed in-house software, including real-time (RT) operating systems (OS), but recently introduced electric control units (ECU) with higher performance. Of these new high performance ECU (HPC), some have been introduced running in-house developed platforms, and some use a supplied software platform based on AUTOSAR Adaptive. Approximately 30 teams work with AUTOSAR Adaptive related design, development, and testing. 

The teams are often organized per application domain (often called "Vehicle Function" internally), and have end-to-end responsibility for their respective functionality. There are also supporting teams, such as a platform team that ensures integration of the vendor supplied platform and provides additional needed platform functionality that is too locally specific to be implemented by the vendor. Another supporting function is the modeling team that provides \texttt{ARXML} modeling for applications and generates service provider skeletons and client proxy implementations for new applications. There are also product managers, architects, system-level testers, and coordination support for teams.

A typical development sequence for vehicle functions starts with the team understanding the requirements through stakeholder discussions. After this the platform is checked to see if there is need for any additions to the services/signals/IO. After this a model can be created and the first versions of the applications can be generated with the vendor-supplied tooling. After this, the team can implement the business logic using \texttt{C++}, test in simulation, on hardware rigs, or in a vehicle. This iteration persists until the desired vehicle behavior is achieved. After this, the vehicle function can be scheduled for release.

\subsubsection{Problem Identification and Motivation}
Through an in-depth workshop series at TRATON R\&D AB, we identified significant pain-points in feature development on a Tier-1-supplied AUTOSAR Adaptive platform. This step was used to collect data for RQ1 and to drive the following steps in the design science process.

We conducted workshops with eight software practitioners, where pain-points were collected. Also, a mapping of the precise development steps of an Adaptive Application from developers' perspectives was constructed. The hand-offs and dependencies between teams were mapped, and the pain-points were connected to the different development steps. We also conducted an initial workshop to develop a reference mapping of workflows based on leader perceptions and the steps described in the development process documentation. The workshops' participants and outcomes were; 1) three senior leaders and managers (project manager, product manager, line manager) mapped the reference journey based on the organization's prescribed processes, 2) two junior feature developers mapped their journey and pain-points, 3) two senior feature developers mapped their journey and pain-points, 4) two platform developers mapped their journey and pain-points in supporting feature development, and 5) two architects were used as a reference group, adding missing parts in the complete journey that was constructed based on the input from the developers.

These subjects were chosen to cover the development steps from the perspectives of all different types of roles involved and to minimize bias from experience. For example, a major pain-point for a junior developer might not seem problematic for a senior developer, and a junior developer might not posses enough understanding to realize a global pain-point that a senior developer can point out.

Through this series of initial workshops, we gathered a rich collection of data on the state of development grounded in the industry context. During our analysis of this data, pain-points were extracted and a taxonomy was defined.The results from the workshop are presented in Section \ref{sec:WS_results}.  

\subsubsection{Objectives of the Solution}
To disentangle the factors influencing the emergence pain-points, our objective was to construct a minimal, independently implemented AUTOSAR Adaptive platform. This artifact served as a diagnostic instrument, enabling controlled reproduction of problematic scenarios and isolating the effects of; 1) Tier-1 implementation, including deviations, extensions, or omissions, and 2) internal development processes, including architectural decisions and tooling assumptions.

Since the possible influences were not distinguishable through observation of the local context alone, due to the entangled nature of the influences, this platform could reset the developers perspectives of what was local, vendor, and AUTOSAR specification specific.

The artifact was designed to support reproducible experimentation on a representative target platform (i.e., it is designed to run on a POSIX OS). Experiments were run on Ubuntu Linux PCs. This research-design choice had two reasons; 1) the process of running our platform on the actual target HPC hardware would have required the platform to conform to local tooling requirements which would have entangled the prototype in local processes--leading to the re-emergence of a confounding factor we wanted to isolate, 2) this allowed for fast experiment-cycles together with the developers.

\subsubsection{Artifact Design and Development}
\begin{figure}
    \centering
    \includegraphics[width=0.9\linewidth]{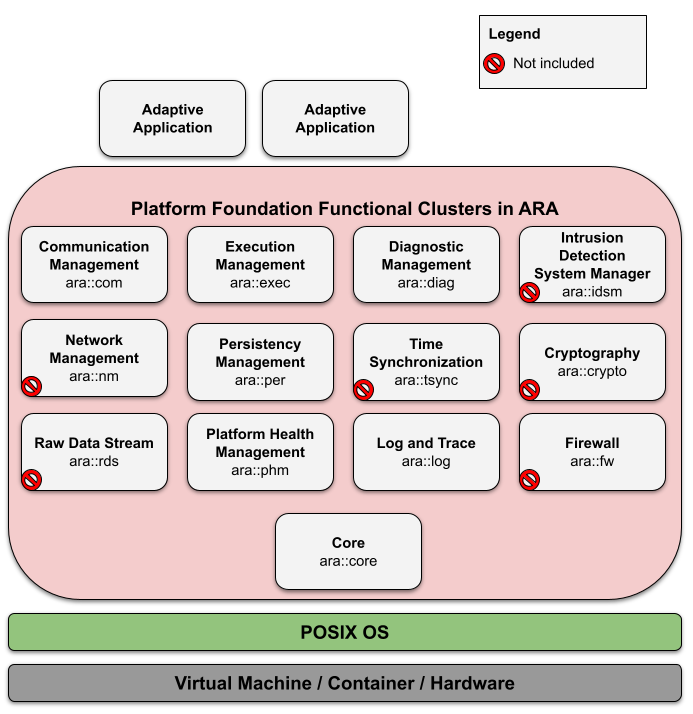}
    \caption{Showing implemented foundational FCs from the AUTOSAR Adaptive platform specifications. Note that the minimal version of \texttt{ara::diag} was added after the static evaluation.}
    \label{fig:FCs_implemented}
\end{figure}

We developed a lightweight AUTOSAR Adaptive platform by selectively implementing core platform services according to the AUTOSAR Adaptive Platform Specification R24-11 \cite{PlatformDesign}\footnote{Available on: https://github.com/SriAbir/minimal-autosar-adaptive}. The implementation minimizes third-party dependencies and allows detailed instrumentation. Core platform Functional Clusters (FCs) are described in \cite{PlatformDesign}, and Figure \ref{fig:FCs_implemented} shows an overview of FCs that were (partially) implemented in the experimental platform.

\textit{Communication Management} handles the previously mentioned service-oriented communication (often over SOME/IP). \textit{Execution Management} handles the application life cycle. \textit{Diagnostic Management} provides diagnostic capabilities, often based on Unified Diagnostic Specification (UDS). \textit{Persistency Management} provides persistent storage capabilities. \textit{Platform Health Management} provides interfaces for the Execution Management to monitor and re-start applications in the event of failure. \textit{Log and Trace} provides interfaces for logging.  The \textit{core} defines a set of common data types used in multiple functional clusters \cite{PlatCore}. The \textit{Intrusion Detection System Manager} \cite{IntrusionMgmt}, \textit{Network Management} \cite{NetwrkMgmt}, \textit{Time synchronization} \cite{TimeSynch}, \textit{Cryptography} \cite{Crypto}, \textit{Raw Data Stream} \cite{RawData}, and \textit{Firewall} \cite{Firewall} clusters were not implemented in the minimal stack.

To allow for implementation in the allotted time for our research, some simplifications had to be made. Another reason for simplification was to avoid implementation details that might reveal confidential material. For these reasons; 1) our platform does not use the \texttt{ARXML} style of modeling specifications \cite{PlatfMethod}. Instead, simplified JSON-based manifests and service descriptions were used to simplify the development, 2) No code generation tools were implemented. Instead, different types of templates were provided for the developers to use, replacing \texttt{Service Proxy} and \texttt{Service Skeleton} generation from a \texttt{Service Interface Definition} \cite{staron2021automotive}, and 3) selective implementation of the bare minimum for developing and running functional requirements of vehicle functions made us choose the functional clusters of ARA, as depicted in Figure \ref{fig:FCs_implemented}. The selection of functional clusters to implement was initially conducted by two of the authors after close consultation of the general AP Platform Design specification \cite{PlatformDesign} and iterated by consulting a development team. Final selection was not done until the Completeness Analysis in the Evaluation step in Section \ref{sec:eval}.

\subsubsection{Demonstration}
The artifact was deployed on Linux Ubuntu as it is a representative POSIX-based operating system, and continuously demonstrated via small applications running on the platform as functionality was added. Although small and simple, these applications had similar behavior as real Adaptive Applications running on the original AUTOSAR platform used by the OEMs supplied by TRATON R\&D AB. 

\subsubsection{Evaluation}
\label{sec:eval}
We applied a multi-step evaluation strategy, including:

\textbf{Conformance analysis:} A systematic comparison of the implemented stack against the AUTOSAR Adaptive standard requirements. We continuously developed runnable applications that used the implemented functional clusters, and confirmed their behavior with the AUTOSAR specifications.

\textbf{Completeness analysis:} One vehicle function candidate was chosen for evaluation. The researchers implemented the application to assess the \textit{completeness} of the stack (the minimal stack needed to be sufficiently large to accommodate the applications). This can be compared with the practice of \textit{static evaluation} described in \cite{Gorschek}. The vehicle function was chosen in consultation with a senior architect as small (meaning no extensive business logic), but representative of the general working of many vehicle functions as it had the characteristics of multiple dependencies both vehicle internally and externally as depicted in Figure \ref{fig:VehicleFunction}.

\textbf{Developer familiarization:} Two developers re-implemented two existing vehicle functions on the minimal platform to become familiar with its workings. The participating developers had engineering degrees, between 5–11 years of software development experience, and between 3–4 years of AUTOSAR-related experience.

\textbf{Developer feedback:} Developers provided detailed comments on the platform and their experiences. In addition to this feedback, a workshop was held to discuss their experiences. The workshop was recorded and transcribed. This session was facilitated as a \textit{focus group in a workshop setting}. As focus groups are a well-established method in software engineering research to elicit and prioritize practitioner perspectives in a structured group format \cite{Kontio}. This method allowed us to elicit and organize the developers' insights and perspectives gained from using the minimal platform. The workshop transcript was thematically analyzed following Braun and Clarke’s coding approach \cite{Braun}. Resulting in a set of workshop themes. 

\textbf{Root cause analysis:} Mapping observed pain-points to their probable origins (standard vs. supplier vs. internal practices) was performed using the 5-Why's method, and these discussions were utilized in the thematic analysis to identify the root causes for the workshop themes. 

The evaluation results provided insight into how the structure and assumptions of the AUTOSAR Adaptive specification manifest in practice and whether specific categories of issues are systemic, implementation-induced, or organizationally situated. The systemic category contained pain-points that the developers could link directly to the insights they gained regarding how the AP is specified. Pain-points in the implementation-induced category were judged to be more closely associated with how the vendor implemented the AUTOSAR requirements. Finally, the organizationally situated pain-points were due to local practices, processes, or team-organization.

\section{Results}
\label{sec:results}
\subsection{Developer pain-points (RQ1)}
The first workshop series identified eleven main potential developer pain-points related to AP development work and its surrounding toolchain, while adhering to local processes. By analyzing developer-supplied digital Post-It notes and workshop transcripts, our taxonomy was constructed from these pain-points. 

\textbf{Platform and Architecture:}
Developers highlighted architectural rigidity and limited observability as core sources of difficulty. 
Strong coupling to supplier-provided frameworks or infrastructure would limit flexibility and slow down iteration. Poor system performance, e.g., for logging, would lead to inefficient trouble-shooting. Vendor supplied platforms could potentially lead to reduced visibility into runtime behavior, which would also hamper efficient development work. This domain reflects the technical boundary conditions that shape developers' control over their work.

\textbf{Knowledge and Information Flow:}
Fragmented documentation and knowledge management practices can lead to the repeated rediscovery of solutions. Information is often stored in isolated repositories or transmitted through informal networks, making it difficult to establish a shared understanding of the intended use of the platform.

Unclear or inconsistent product requirements were also mentioned as a possible contributing factor to uncertainty about purpose and verification criteria. This would result in an increase in testing time. This domain emphasizes the cognitive overhead caused by the lack or inaccessibility of organizational knowledge.

\textbf{Process and Workflow:}
Developers described repetitive manual steps, such as flashing, testing, and integration, that were only partially automated as possible pain-points. Unclear mandatory procedures and inconsistent process ownership created uncertainty about how to progress the work and who was responsible for approvals. A fragile release chain was also considered problematic, with long rebuild times and limited transparency across the CI/CD stages as a possible result. These issues point to the absence of smooth and predictable workflows for feature integration and validation.

\textbf{Tools and Infrastructure:}
If the surrounding tool landscape was fragmented and difficult to access specifically for high-performance computing environments lacking a standardized toolchain or consistent access management would lead to developer experienced pain-points. This would create wait-time for permissions or licenses and encouraged the emergence of locally developed tools and scripts that would further increase divergence. Potential observability gaps, such as missing telemetry, would compound the sense of infrastructural fragility.

\textbf{Collaboration and Coordination:}
Cross-organizational communication was cited as a possible source of friction. Differences in processes, terminology, and tool usage between teams and brands would hamper coordination. Unclear ownership of interfaces and requirements would potentially add to misalignment, causing redundant work and late integration problems.  This domain captures the socio-technical coordination cost introduced by distributed development.

\textbf{Work Execution and Scope Management:}
Developers reported difficulties maintaining focus due to fragmented workdays, frequent meetings, and multi-branch maintenance as potential pain-points. 

In addition, if many features were scoped too broadly, reducing opportunities for incremental testing and early feedback this would also be troublesome. Although sometimes, scoping smaller features was not always possible. These factors would collectively slow down learning and increase the risk of rework.  This domain reflects how local work practices and planning decisions interact with technical constraints to influence productivity.

\begin{figure}
    \centering
    \includegraphics[width=0.9\linewidth]{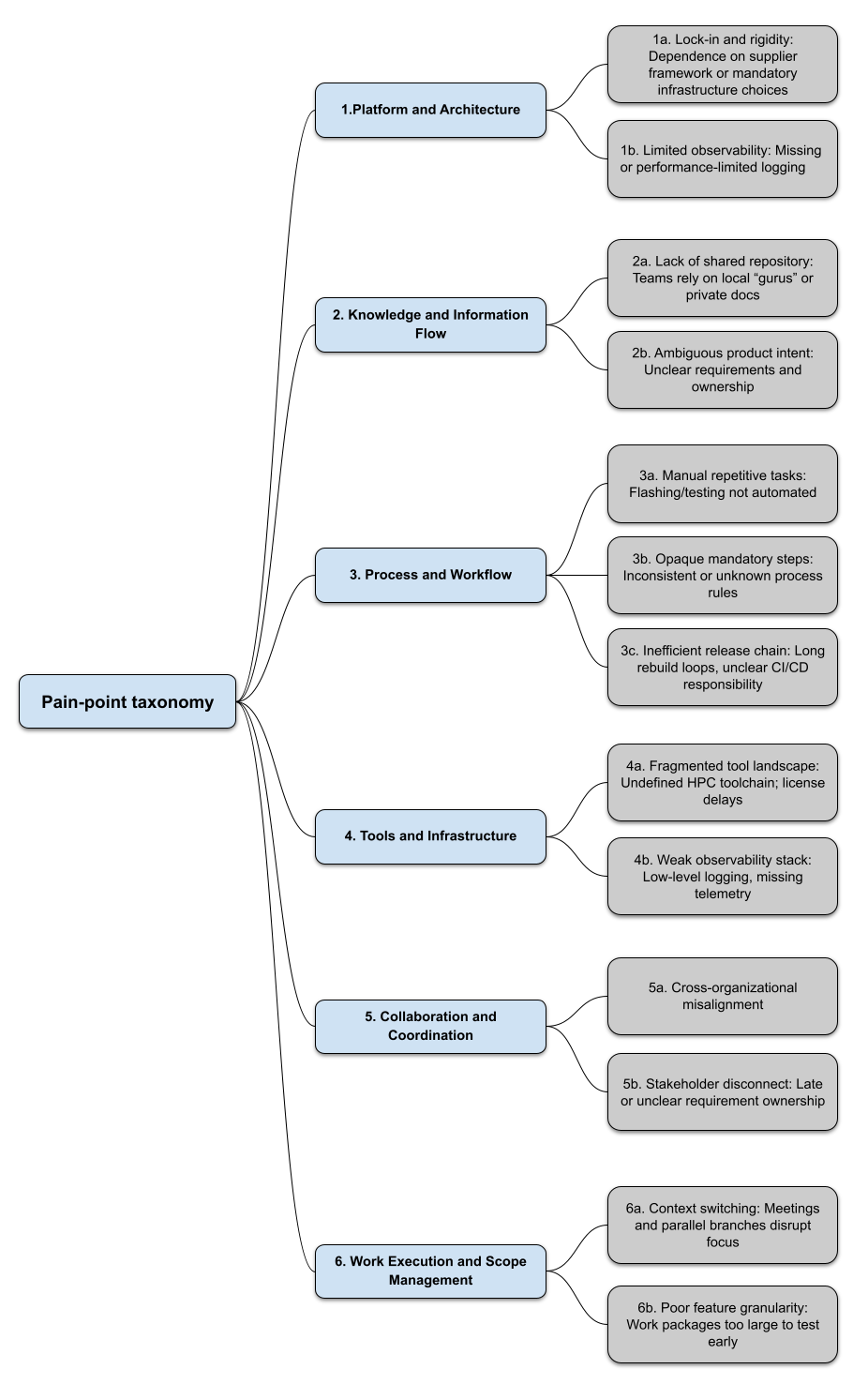}
    \caption{Taxonomy of pain-points from the initial workshop series.}
    \label{fig:pp_taxonomy}
\end{figure}

\textbf{Summary:}
An overview of the taxonomy based on the data collected in the initial workshop series is shown in Figure \ref{fig:pp_taxonomy}. In order to verify the categories and ascertain the root cause of the pain-points, and how closely related they are to the AP specification, we evaluate our minimal AP stack in an effort to isolate the influence of AUTOSAR, vendor, and local complexities. 

\subsection{Static evaluation}
Among the approximately 30 teams working on different aspects of AUTOSAR Adaptive development, a team responsible for a set of vehicle functions connected to the connectivity-HPC was selected as a collaborator for static evaluation. The reasoning behind this choice was that one function for which the team was responsible had just the right size and complexity to constitute a realistic example of function development, while at the same time being reproducible on the new AUTOSAR Adaptive stack within a manageable time frame. 
 
\begin{figure}
    \centering
    \includegraphics[width=0.95\linewidth]{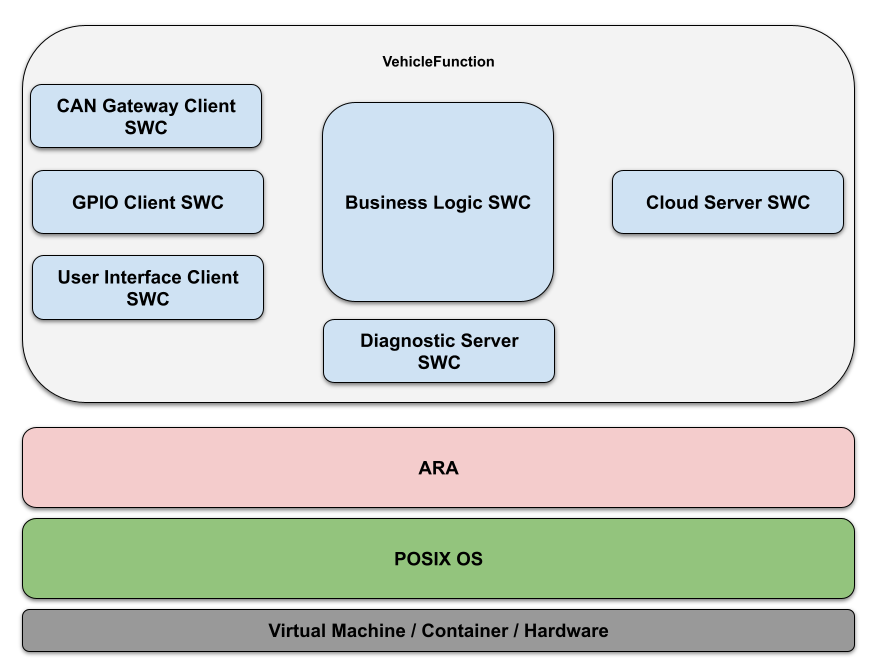}
    \caption{Depiction of the major parts of the VehicleFunction application.}
    \label{fig:VehicleFunction}
\end{figure}

The actual application chosen (called VehicleFunction in the rest of this paper) cannot be described in detail due to a Non-Disclosure Agreement with our industrial partner. Still, a sketch of the implementation is depicted in Figure \ref{fig:VehicleFunction}. This application was selected as a representative for evaluation because it supports service discovery, request/response, publish/subscribe, timers, and state machines in multiple applications. It also validates error handling and degraded modes.

Since the application is dependent on other implementations and external interfaces, certain simplifications were necessary. Dependencies were handled with stubs that had reduced behavior because the alternative would have been to replicate the entire system in question (a multi-year endeavor). The functional compliance between the simplified and original vehicle function was checked with the responsible development team and was deemed similar enough to ensure the completeness and conformance analysis of the platform.

During the static evaluation, the researchers implemented the application to assess the stack before the developer evaluation. We found that two features of the minimal stack were missing, which we added before the developers' evaluation; 1) \textit{Diagnostics:} A small DoIP server (TCP/UDP according to ISO 13400) and ara::diag shim (Diagnostic Management \cite{Diag}) that forwards a handful of UDS services, and 2) \textit{In vehicle communication:} A small SocketCAN daemon that subscribes to VehicleFunction events and transmits the corresponding CAN frames.

To avoid revealing internal company implementation details, this addition was split into a public (available on GitHub) and a non-public part. It was also decided to keep stubs for cloud service communication, position information, and user interface connections non-public.

\begin{figure}
    \centering
    \includegraphics[width=0.99\linewidth]{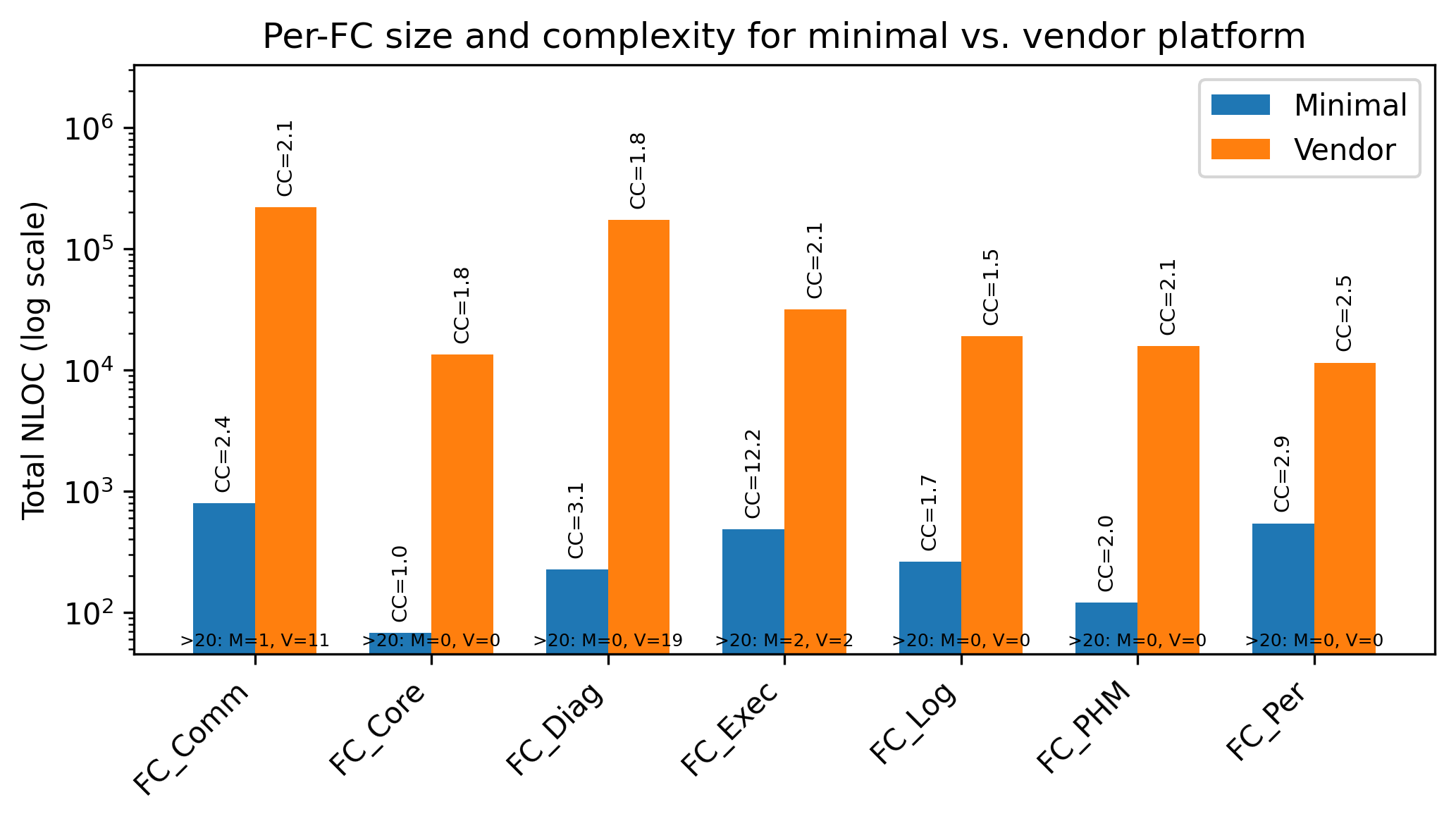}
    \caption{Showing a comparison between the minimal and vendor supplied platforms. Lines of code in logarithmic scale to allow visualization in the same image. CC label is average cyclomatic complexity per function. Number of functions with CC \textgreater{} 20 shown for M=minimal and V=vendor. The logarithmic scale is used to visualize large NLOC differences. For example, the figure illustrates that while Execution Management exhibits inherent complexity in both stacks, vendor implementations accumulate substantially more complexity across multiple functional clusters.}
    \label{fig:fc_comparison}
\end{figure}

We performed a lightweight static analysis of FCs in both the minimal and vendor-supplied AUTOSAR platforms, focusing on cyclomatic complexity (CC) and non-commented lines of code (NLOC). \textbf{In the minimal platform}, most FCs exhibited low average CC values (1–3) and no functions exceeding CC \> 20. The Execution Management FC had a higher average CC ($\approx$ 12) with two high-complexity functions, consistent with the inherent coordination logic mandated by the AP specification. \textbf{The vendor platform} showed higher complexity across several FCs, with Communication, Diagnostics, Persistency, and Logging all containing functions above CC \> 20 and larger total NLOC. This indicates that while some complexity is inherent in the AP architecture (e.g., lifecycle management), a portion of the complexity experienced by developers originates from vendor-specific implementations and the requirements on automotive software requirements in general. See Figure \ref{fig:fc_comparison} for details.

\subsection{Developer evaluation (RQ2, RQ3)}
The final evaluation of the minimal AUTOSAR Adaptive artifact was conducted during a workshop with two developers. The developers were chosen from teams separate from the team that supported the static evaluation. 

\begin{table*}[!htbp]
    \centering
    \caption{Results from the developer workshop during the evaluation phase of our Design Science Research project.}
    \begin{tabular}{|p{3cm}|p{1.5cm}|p{5cm}|p{5cm}|p{2cm}|}
        \hline
        \textbf{Evaluation themes} & \textbf{Related to AP specification} & \textbf{Root cause analysis} & \textbf{Possible negative organizational impacts} & \textbf{Taxonomy reference}\\
        \hline
         Vendor lock-in & Yes & The Adaptive Platform is over-specified in some areas (e.g., determinism, service discovery) but vague in others, forcing vendor-specific interpretations and tooling dependencies (e.g., AMSR-SDK). & Application code tightly coupled to vendor-specific libraries and service definitions; limited portability and increased maintenance cost . & 1a , 1b\\
        \hline
        Over-engineered service discovery and callbacks & Yes & SOME/IP-based service discovery and timing constraints create multiple callbacks and timers. Complexity stems from enforcing determinism and runtime flexibility simultaneously. & Unreadable asynchronous code scattered across applications; duplicated (cloned) callback and timer logic instead of a reusable library abstraction. & 1a, 1b\\
        \hline
        Hard to learn mental model of lifecycle & Partly & The AUTOSAR app lifecycle (startup, shutdown, execution management) is complex and poorly documented, making it difficult for new developers to build a coherent mental model. & High onboarding cost and defensive coding patterns; reliance on framework “magic” and avoidance of deeper refactoring. & 1b, 2a, 2b\\
        \hline
        Slow build / heavy code generation & Partly & The platform rebuilds serialization and deserialization code for all PGNs and services instead of modularizing generation. & Long build times discourage experimentation and iteration; redundant code generation leads to inefficiency. & 3c, 4a, 5a \\
        \hline
        Opaque ARXML and generated JSON configs & Yes & Tooling produces massive configuration files (e.g., 26k-line SOME/IP configs) that are unreadable and runtime-parsed even when known at compile time. & Fragile configuration management, risk of runtime mismatches, and difficulty debugging service definition issues.  & 1a, 3b\\
        \hline
        Unclear value of AUTOSAR (“what does it bring?”) & No (organizational) & Developers perceive AUTOSAR as a constraint rather than an enabler. The architectural rationale is poorly communicated relative to simpler OS-level service management. & Cultural and process debt; potential emergence of parallel “shadow stacks” to bypass the official platform. & 3b, 5b, 6b\\
        \hline
        Complexity added by internal modeling & No (local practice and vendor implementation) & Internal modeling restrictions (e.g., application access requests, ARXML gating) add unnecessary complexity, blurring the line between vendor and AUTOSAR behavior. & Persistent confusion about responsibility boundaries; added maintenance cost for local extensions. & 2b, 3b, 5a, 5b\\
        \hline
        Over-specification and under-guidance & Yes & The AP specification attempts to cover all use cases but lacks developer-facing guidance, resulting in inconsistent interpretations and steep learning curves. & Documentation debt and fragmentation; need for internal simplifications or “teaching stacks” like the minimal platform. & 1a, 1b, 2b\\
         \hline
    \end{tabular}
    
    \label{tab:workshop_results}
\end{table*}

Before the workshop, the developers re-implemented two existing vehicle functions to familiarize themselves with the minimal platform. Both of these functions were representative of vehicle functions performing business logic and having multiple dependencies. One of the functions was similar to the VehicleFunction (see Figure \ref{fig:VehicleFunction}) in structure and enabled users to interact with vehicle-internal functionality and vehicle-external connected devices. In order to target a wider range of functionality types, the other function that was chosen had more of a supporting role in the system, creating an isolated local environment in the communicator-HPC to run early prototype testing of applications.

Similar to static evaluation, the exact functions of production-grade vehicles could not be recreated. However, the developers assessed that their experience familiarizing themselves with the experimental platform was sufficient for them to compare it with working on the existing platform.

As they worked, the developers documented their experiences developing on the prototype platform in a journal . At the end of the familiarization period, they were asked to summarize their experiences and relate them to the pain-points in a workshop setting.

\subsubsection{Workshop results}
\label{sec:WS_results}
The group distinguishes between problems intrinsic to AUTOSAR Adaptive (service discovery, determinism, ARXML heaviness) and problems that emerge from vendor or local adaptations (tooling lock-in, model restrictions). A recurring meta-theme is that complexity has shifted from integration to comprehension: developers must either reverse-engineer vendor behavior or re-implement minimal abstractions to regain control. The main findings of the workshop are shown in Table \ref{tab:workshop_results}. 

The developers were asked to relay their experiences comparing their work with the minimal prototype to their experience with the currently used platform. The results are shown in Table \ref{tab:dev_exp}.

\begin{table*}[!htbp]
    \centering
    \caption{Showing the direct quotes from developers when comparing the prototype platform with the current in-use platform.}
    \begin{tabular}{|p{9cm}|p{9cm}|}
        \hline
         \textbf{Current Platform} & \textbf{Prototype Platform}\\
         \hline
         "Hard to learn for new users." & "Easier to learn for new users." \\
        \hline
         "Slower build." & "Faster build." \\
         \hline
         "More control over app details." & "Hides some important details."\\
         \hline
         "Easier to work with when you get familiar." & "Not sure it is gonna be easier to cope with difficulties you normally face with AUTOSAR and C++ projects."\\
         \hline
         "Feels overengineered. The whole process of discovering a service involves setting up multiple callbacks and timers, etc" & "Goes to the point. Allows the developer to focus on the core logic instead of boilerplate." \\
         \hline
         "Heavily dependent on ARXML files that are very hard for humans to read." & "Service definitions and data types are defined in headers." \\
         \hline
         "Generated files (JSON-files containing service info) are massive." & "vsomeip/local.json is quite minimal. Of course this is a local example, but it contains all the information needed for service discovery" \\
         \hline
         "The model generates serialization code automatically" & "Currently, serialization must be implemented manually; this is a problem as it needs to be done carefully." \\
         \hline
    \end{tabular}
    
    \label{tab:dev_exp}
\end{table*}

\section{Discussion}
\label{sec:discussion}
Our findings show that many of the perceived pain-points in working with the AUTOSAR Adaptive Platform (AP) do not stem from the specification alone but from the interaction between the standard, supplier implementations, and local organizational practices. The complexity of the AP is thus both \textit{architectural} and \textit{epistemic}: it emerges from how knowledge about the system is distributed and mediated through tools, models, and specifications.

The minimal stack was not intended as a production alternative nor as a complete AUTOSAR implementation. Its purpose was diagnostic: to isolate specification-mandated architectural mechanisms (e.g., lifecycle management, service discovery) from vendor-specific extensions and local tooling constraints. By simplifying modeling (e.g., replacing ARXML with JSON) and removing SDK dependencies, we intentionally reduced accidental complexity to observe which difficulties persisted. Those that remained are strong candidates for specification-inherent complexity.

\subsection{RQ1: Developer pain-points}
The workshops revealed that developers experience the current platform as rigid, opaque, and slow to iterate on. While the Adaptive Platform aims to provide abstraction and safety through standardization, developers instead encounter high coupling to vendor-specific implementations and large generated artifacts (e.g., ARXML, SOME/IP JSONs). As a result, everyday tasks, such as configuring communication or debugging runtime behavior, become disproportionately effortful. These findings mirror observations in previous studies on AUTOSAR complexity and coupling \cite{Durisic}. However, it was clear from our results that some pain-points are inherent in the AP specification from AUTOSAR and would likely always persist.

For example, developers reported uncertainty regarding when shutdown hooks are guaranteed to execute and how restart semantics interact with persistent state. This uncertainty led to defensive patterns such as redundant state validation and additional error-guarding logic, increasing code complexity. In practice, a simple client-server interaction required multiple asynchronous callbacks, timers, and state checks, even when the communication topology was static. This scaffolding often overshadowed business logic and was frequently re-implemented across applications.

\subsection{RQ2: Root causes}
Through the comparative evaluation of the minimal AUTOSAR stack, developers were able to distinguish which problems originate in the specification itself (presented in Table \ref{tab:workshop_results}. The key insight is that the AP specification attempts to cover a wide set of use cases but provides little developer-facing guidance on how to realize them. This is aligned with our static analysis results, where only Execution Management is inherently complex in the minimal platform, whereas the vendor platform accumulates additional complexity across several FCs. The under-specification leads to divergent vendor interpretations—particularly around service discovery and execution management—which in turn propagate to OEM-level complexity. Our result show that developers must navigate a “stack of interpretations,” where the boundary between standard and implementation is blurred. However, it is unclear how a general specification could be achieved without similar effects. 

Importantly, not all identified pain-points are specific to AUTOSAR Adaptive. We distinguish between three causes: 1) specification-inherent issues (e.g., service discovery callbacks, lifecycle management complexity, ARXML-heavy configuration), 2) vendor-induced complexity (e.g., tooling lock-in, generated artifact size, SDK coupling), and 3) generic large-scale software issues (e.g., fragmented documentation, coordination overhead).

Our contribution is not to claim that AUTOSAR uniquely causes all observed difficulties, but to demonstrate which difficulties are amplified or structurally shaped by specific architectural choices in the Adaptive Platform.

\subsection{RQ3: Organizational implications}
Several of the pain-points identified have a negative organizational impact. The reliance on vendor-specific tooling and generated artifacts introduces lock-in and hinders flexibility. The extensive callback logic and service-discovery scaffolding required to achieve determinism in SOME/IP-based communication incurs hidden maintenance costs. In addition, the limited transparency of lifecycle management encourages defensive coding and reduces changeability. These forms of architectural debt need to be actively managed to avoid negative organizational implications.

\subsection{Reflections on specification versus implementation}
The experimental minimal platform provided a unique lens for isolating the role of the AUTOSAR specification itself. Developers appreciated the simplicity of the minimal stack and the clarity it provided for understanding how applications start, stop, and communicate. This suggests that the complexity of the specification could be mitigated through pedagogical tools or “teaching stacks” that make life-cycle behavior explicit. Such artifacts could help organizations build shared mental models of the platform without relying solely on vendor black boxes. 

However, even though the minimal stack "goes to the point" as one developer put it, it's not clear that this sense of ease will persist in a full-scale platform. Specification details will likely be obscured as the platform grows, and minimize the positive effects experienced by the developers. Building large-scale embedded software systems with many dependent software clusters is an essentially complex endeavor \cite{Antinyan} and there will be trade-offs between the benefits and downsides of using a centrally specified software platform. 

\subsection{Implications for industrial practice}
The results indicate that the successful adoption of the AUTOSAR Adaptive Platform depends not only on conformance to the standard but also on transparency in its interpretation. Organizations should therefore; 1) ask for clearer documentation and diagnostics from suppliers to bridge the gap between standard and implementation, 2) establish internal competence-building initiatives using simplified and instrumented stacks, and 3) treat configuration and model management as first-class development concerns rather than toolchain artifacts.

In this way, industry actors can take advantage of the benefits of standardization while easing the effects of the essential complexity of large scale embedded system development by not adding accidental complexity \cite{Antinyan}. 

\subsection{Validity Considerations}

The validity of this study is influenced by both methodological and contextual factors. We discuss construct, internal, and external aspects in turn.

\textbf{Construct validity:}  
To ensure that the collected data accurately reflect developer experience with the AUTOSAR Adaptive Platform (AP), we triangulated multiple sources: mapping of development-steps, workshop discussions, and hands-on experimentation on a minimal AP stack. Pain points were elicited using open-ended prompts and iteratively constructed in group settings to reduce researcher interpretation bias. However, since the notion of a “pain-point” is partly subjective, results may still depend on participants’ interpretation within their organizational setting.

\textbf{Internal validity:}  
Pain-point identification was conducted collaboratively between two of the authors. The taxonomy was individually developed but checked by the other authors of the paper. Linking the taxonomy domains to specific root causes was done collaboratively by the workshop participants (facilitated by two of the authors) to minimize individual bias. The design-science artifact served as a control mechanism to separate the effects of the specification from those of implementation. Nonetheless, the presence of residual confounding factors remains possible since the experimental stack does not implement the full AUTOSAR standard, and simplifications may have affected the experienced comparison between the two platforms.

\textbf{External validity:}  
The study was conducted within one industrial group and involves a limited number of developers who are familiar with a single vendor’s implementation. While the identified pain-point domains in the taxonomy are likely transferable to similar large-scale embedded software contexts, their relative weight may differ across organizations or different AUTOSAR Adaptive implementations. However, the results should generally hold across similar contexts. We tried to mitigate threats to external validity by including developers with a broad range of experience in both the workshop series (RQ1) and prototype evaluation (RQ2/3). 

\section{Conclusions}
\label{sec:conclusion}
The goal of this study was to investigate the pain-points developers experience with AUTOSAR and whether they originate from the platform, its implementation, or its usage. We were interested in collecting pain-points in a local context, followed by tracing their sources and possible root causes to the AP architecture. We did this by isolating the analysis from the effects of local tooling, implementations, and practices via an experimental (minimal) AP implementation that was evaluated by developers from the local context. Developing an application on the prototype platform enabled the developers to compare their experiences from day-to-day work with a more idealized AP-setting.

Our findings are that the specification of the AP may lead to; 1) \textbf{vendor lock-in}, that systems developed on the platform may suffer from over-engineered service discovery and callbacks, 2) have a \textbf{hard to learn mental model} of application life-cycle, 3) \textbf{suffer from slow builds} and heavy code generation, 4) need \textbf{large and hard to read configuration files}, and 5) \textbf{reliance on requirement specifications} that are over-specified while lending too little guidance. 

The conclusions from our DSR study are that there are complexities introduced by the AUTOSAR specifications. The main contributors are the service discovery mechanisms for managing software component dependencies. However, it is also clear from the study that large-scale embedded software development is by no means trivial. Complexities are an emergent property of the distributed nature of the products being developed. To some extent, the impact can be mitigated by education and tooling, but introducing AP technologies will require developer-time spent on building understanding and mental models for application life-cycles before development can become efficient and effective.

The Adaptive Platform introduces essential complexity through architectural mechanisms required for distributed, service-oriented vehicle systems. However, our findings suggest that specification under-guidance combined with vendor interpretation amplifies this complexity beyond what is strictly necessary.

In future work, we aim to extend this study with comparisons of competing frameworks such as ROS2 \cite{Hong} and Eclipse S-CORE \cite{S_CORE} to provide further illumination of the complexities inherent in the AUTOSAR Adaptive specifications. We also plan to extend the data collection with quantitative data in order to investigate performance-related findings from this study.

\end{document}